\documentstyle[]{article}
\begin{document}
\begin{center}
{ \huge \bf
Mechanical and dielectric response of microcomposites of the type: ferroelastic-dielectric}
\end{center}
\vskip3cm
\begin{center}
{\large  O. Hudak\footnote{e-mail:hudako@mail.pvt.sk} \\
\vskip1cm
Institut of Experimental Physics,University of Vienna, Boltzmanngasse 5, A - 1090 Wien 
\\
\vskip1cm
W. Schranz\\
\vskip1cm
Institut of Experimental Physics,University of Vienna, Boltzmanngasse 5, A - 1090 Wien 
\vskip1cm
}
\end{center}

\begin{center}
\vskip1cm
	Matej Hudak
\vskip1cm
\end{center}

\begin{center}
	Stierova 23, Kosice, SK-040 23 Slovakia
\end{center}

\newpage
\section*{Abstract}

Dynamic dielectric and mechanical responses of the 
ferroelastic-dielectric microcomposite were studied in this paper. The mechanical
inclusion-matrix interactions have influence on the mechanical moduli of the composite. We have described a mechanical response of the composite which consists of the material M and of the other material I with dispersion of mechanic moduli of particles due to near sphere shape and due to their anisotropy. Larger dispersion of mechanical moduli of the second material leads to softening and the stiffness has lower values for small concentrations. The inclusions of the dielectric material in the ferroelastic matrix has the effect also on the dielectric response of the microcomposite. We have found under which conditions the inclusions have the effect of enhancement of the dielectric response.
Results of low frequency dielectric constant and dielectric loss in
orthorhombic $ Al_{2}(WO_{4})_{3}$ show for polycrystalline material,
where voids may play the role of dielectric material, linear
increasing with increasing hydrostatic pressure. This is qualitatively in
agreement with our theory. We are explaining the presence of the hysteresis phenomena observed in these experiments by amorphization which changes free volume and its effective mechanical properties.

\newpage
\section{Introduction}

The problem of effective shear and bulk moduli, of an effective Poison's ratio and of an effective dielectric response in microcomposites of the ferroelastic-dielectric type is a very interesting one. 
The static and dynamic dielectric response of ferroelectric-dielectric and the static response of ferroelastic-dielectric type microcomposites
was studied recently \cite{HS} - \cite{OH2}.  A coupling of the elastic
strain to the electric polarization leads to the dielectric
response of a ferroelastic type material,  see in
\cite{jona} - \cite{LG}. A high anisotropy of single crystal properties induces an enhancement in the macroscopic piezoelectric response for polycrystalline materials like $BaTiO_{3}$ and $PZN-PT$, \cite{GCL}. The authors of the cited paper do not consider pores in the material. However there is always a free volume in a polycrystalline material which is formed by these pores. We can consider polycrystalline materials as a microcomposite in which one type of material is a crystalline material and the other one are pores. The authors of the mentioned paper determined macroscopic properties of a polycrystal from the resulting average strain tensor. They compare results of their modelling with experimental data. While they have found that texture and the corresponding grain size enhance piezoelectric response, one may expect that also the anisotropy of clusters increases when the concentration of one component of the microcomposite increases leads to similar effects, when the clusters become very large. The clusters are not isotropic, they are anisotropic. Thus we may expect that in microcomposites of the ferroelastic-dielectric type increase of the concentration of dielectric material leads to the enhancement of the dielectric response. 
Materials of the perovskite type ($LaAlO_{3}$, $CaAlO_{3}$, $SrAlO_{3}$, $BaTiO_{3}$, $PbNiO_{3}$, $Pb(Zr,Ti)O_{3}$, ...)
undergo a phase transition from a cubic phase to a phase with lower
symmetry at some critical temperature. 
The dynamic mechanic response, and the dynamic dielectric response of such materials which is due to coupling between the elastic strain tensor and the polarization is thus an interesting problem.

We have studied recently the static response of microcomposites of the type mentioned above \cite{HS}, and discussed qualitatively the results of the model for pressure dependence of a dielectric constant in the $Al_{2}(WO)_{4}$ polycrystalline material, in which a second material were porous in between crystallites.
Let us note that the $Al_{2}(BO)_{4}$ type materials are used in many applications like fuel cell electrolytes, gas sensors, laser materials, composites with tunable thermal expansion, see in \cite{MVATG} and in references therein.
Mechanical analysis measurements are done usually at low
frequencies (0.1Hz - 10 Hz), however measurements at higher frequencies
are done also, \cite{STKSS} - \cite{KSSHSS}. Elastic response function (compliance)
shows Cole-Cole diagrams of the circular and of the non-circular behavior for 
these materials in their crystalline and in their ceramic form \cite{HRS}.  Multirelaxation phenomena exist in these 
materials under some conditions.

It is known that higher electric and mechanical loading
leads to a nonlinear behavior in ferroelectric and ferroelastic
ceramics \cite{E}. At higher temperatures however there is present no
ferroelastic phase, there is present a paraelastic phase. Coupling of the elastic strain and electric
polarization does however exist at these temperatures too. Mechanical and dielectric response 
depend then on mechanical forces acting on the microcomposite of the
ferroelastic-dielectric type at these higher temperatures. Constraints
due to neighboring material lead to a 
transition from the paraelastic to  a ferroelastic phase \cite{JCD}. We assume
in this paper that there are small mechanical fields of the order
$10^{-3}$ which are acting on the microcomposite, and that particles in the microcomposite have a diameter of the order of
1$\mu$m, see also \cite{HS}. For large concentrations of
the dielectric material the clusters of the ferroelastic phase 
are small and finite. At the percolation transition there are very large
clusters of ferroelectric and dielectric materials. It is not known
 how they are responding to the external mechanical fields mainly due to the fact that the form of the clusters is irregular. 
To study the region of the percolation transition in the microcomposites of the ferroelastic-dielectric type we need to understand the behavior of elastic moduli of the microcomposite at this region.

In a ferroelastic phase long-range anisotropic forces may appear
\cite{LSRSB} in the region when large inclusions are present. In our paper we will discuss properties of the microcomposite
above the critical temperature. Thus the long-range forces can be
neglected. To determine the behavior of elastic moduli of the microcomposite we assume that the grains of both phases are more-less of the spherical shape. The shape of the ferroelastic material particles is assumed to be spherical and characterized by the same diameter, it dependens on the method of preparation of the microcomposite. The shape of dielectric particles is assumed to be more-less spherical and we assume further that there exists a distribution of their radii and of their shapes which may be characterized by a homogeneous distribution function as concerning mechanical properties. Analysis of the elastic moduli of heterogeneous materials was done in \cite{B}. We will use this method to analyze our system of the ferroelastic-dielectric  microcomposite. We will find mean-field equations for the effective shear and bulk moduli of the composite, and for the effective Poisson's ratio.

The aim of this paper is to study the dielectric and mechanical response of microcomposites of the ferrolastic -
dielectric type. The mean-field equations for the effective shear and bulk moduli of the microcomposite together with the mean field equation for the dielectric susceptibility enable us to study these responses.

\section{Model of ferroelastic-dielectric microcomposites}

Let us describe a model for ferroelastic-dielectric microcomposites. The polarization, a primary order parameter in ferroelectrics, is in ferroelastics  a secondary order parameter. Primary
order parameter in these later materials is the corresponding component of the elastic
strain tensor (or a combination of components of the elastic strain tensor).  The ferroelectric
state is present in such a material due to a coupling between the
elastic strain tensor and the polarization. Changing
the concentration of two types of particles in the
microcomposite the response to external fields changes. This holds for
dielectric response and for mechanical responses. As we noted for simplicity we consider all particles in the
microcomposite of the same diameter d for the ferroelastic material and of nearly the same diameter and of nearly spherical shape for dielectric material, which is a realistic approximation. This enables us to study microcomposites in which dielectric particles may be also porous, and thus polycrystaline materials.
The concentration at which an infinite cluster of dielectric  particles 
forms is the percolation transition concentration.  Increasing
further the concentration of the dielectric material this material becomes the matrix
material. Thus the interplay between the
percolation transition and the ferroelastic phase
transition appears. 

\section{Elastic moduli of the microcomposite}

We will assume that the microcomposite consists of particles - spheres of the material M, we will use the index N for the material constants of this material,  and of particles - almost spheres which have different "diameters" of the other material. We will assume further that there are (N-1) particles of the second type. Mechanically we thus have a composite from N types of particles. Let us assume that the concentration of the particles of the M material is 1-x, and that the $i=1, ..., N-1$ types of particles of the second material I have the same "concentration" $x_{i}= \frac{x}{N-1}$ . Thus the concentration of the second material is x. We can imagine this composite as if it is to consisted of regular grains of the material M and of less regular, even almost irregular, grains of the second material which are almost spheres and which have some distribution of diameters. The size distribution of grains of the second material is assumed to be homogeneous. Then mechanically we have a material of the volume V with the N phases. The volume concentration is $ \frac{V_{i}}{V} $ for the i-th particle. We assume that the particles of the second material have the same volume $ \frac{Vx}{N-1} $ . Derivation of effective moduli using the method according to \cite{B} leads to the effective shear modulus $G^{*}$ and to the effective bulk modulus $K^{*}$ and to the effective Poisson´s ratio $\nu^{*}$. We consider, as in \cite{B},  a large cube with edges parallel to the coordinate axes (x,y,z). We further assume that a uniform shear stress acting to the surface of the cube and a uniform hydrostatic pressure acting on the surface of the cube lead to a strain and a volume contraction. Note that for N large the particles of the second material in the material M will appear as inclusions in a matrix consisting of that N phase. However in \cite{B} it is used the composite as a matrix, and thus he includes interactions of these particles of the second material in the material M as the concentration of these particles increases. This then leads to effective elastic moduli. 

Let us now describe equations for the effective elastic moduli in the limit of a large number of particles of the second material.
The effective shear modulus $G^{*}$ is found from the equation:

\begin{equation}
\label{1.}
\frac{1}{G^{*}} = \frac{1}{G_{N}} + x \int^{1}_{0} dr (1 - \frac{G(r)}{G_{N}}) \frac{1}{G^{*} + \beta^{*}(G(r) - G^{*})}
\end{equation}

where:

\begin{equation}
\label{2.}
\beta* = \frac{2(4 - 5 \nu^{*})}{15(1 - \nu^{*})}
\end{equation}

and $ \nu^{*}$ is the effective Poisson´s ratio of the composite material.

The effective bulk modulus $K^{*}$ is found from the equation:

\begin{equation}
\label{3.}
\frac{1}{K^{*}} = \frac{1}{K_{N}} + x \int^{1}_{0} dr (1 - \frac{K(r)}{K_{N}}) \frac{1}{K^{*} + \alpha^{*}(K(r) - K^{*})}
\end{equation}

where:

\begin{equation}
\label{4.}
\alpha^{*} = \frac{1 +  \nu^{*}}{3(1 - \nu^{*})}
\end{equation}

and $ \nu^{*}$ ,  the effective Poisson´s ratio of the composite material, is given by:

\begin{equation}
\label{5.}
\nu^{*} = \frac{3 K^{*} -  2 G^{*}}{6K^{*} + 2 G^{*}}
\end{equation}

Thus equations (\ref{1.}) - (\ref{5.}) give the dependence of $ G^{*} $  and $ K^{*} $ on the elastic constants $G_{N}$ and $K_{N}$, and $G(r)$ and $K(r)$, the distribution functions of shear and bulk moduli for the second material. Here r denotes different types of particles of the second material.

Let us assume that the distribution of the elastic moduli for particles of the second material is linear:

\begin{equation}
\label{6.}
G(r) = r G_{1} + (1 - r) G_{0}
\end{equation}

and:

\begin{equation}
\label{6´.}
K(r) = r K_{1} + (1 - r) K_{0}
\end{equation}

where $G_{0}$ and $G_{1}$ are two values of the shear modulus G for the second material which correspond to the index 0 and 1 of the particles, and where $K_{0}$ and $K_{1}$ are two values of the bulk modulus K for the second material which correspond to the index 0 and 1 of the particles. Let us note that the distribution may be in general different from the linear, however the linear approximation is expected to be good for the case in which values for the moduli for $r =  0$ and for $r = 1$ are not different too much. Note that for larger difference $ G_{1} - G_{0}$ , i.e. for the second material in which rigidity of particles changes with the change of the shape and the size, e.g. due to presence of texture, cracks or other types of defects.
Then we obtain equations (\ref{1.}) - (\ref{5.})in the following form.

The effective shear modulus $G^{*}$ is found from the equation:

\begin{equation}
\label{7.}
\frac{1}{G^{*}} = \frac{1}{G_{N}} + \frac{x}{G_{N} \beta^{*}} [\frac{\beta^{*}(G^{*} - G_{N}) - G^{*}}{\beta^{*} (G_{0} - G_{1})} \ln (\frac{G^{*} + \beta^{*}(G_{1} - G^{*})}{G^{*} + \beta^{*}(G_{0} - G^{*})}) - 1]
\end{equation}

where $ \beta^{*}$ is given (\ref{2.})
and $ \nu^{*}$ is again the Poisson´s ratio of the composite material.

The effective bulk modulus $K^{*}$ is found from the equation:

\begin{equation}
\label{9.}
\frac{1}{K^{*}} = \frac{1}{K_{N}} + \frac{x}{K_{N}\alpha^{*}} [\frac{\alpha^{*} (K^{*} - K_{N}) - K^{*}}{\alpha^{*}(K_{0} - K_{1})} \ln (\frac{K^{*} + \alpha^{*}(K_{1} - K^{*})}{K^{*} + \alpha^{*}(K_{0} - K^{*})}) - 1]
\end{equation}

where $ \alpha^{*}$ is given in (\ref{4.}).

From the equations (\ref{7.}) - (\ref{9.})it follows that such textures, cracks and defects which induce larger difference of the elastic moduli G and K lead in the first approximation of the inverse of the absolute value of the difference $ G_{1} - G_{0}$ and of the difference $ K_{1} - K_{0}$ to the microcomposite in which:

\begin{equation}
\label{9... }
\frac{1}{K^{*}} \approx \frac{1}{K_{N}} (1 - \frac{x}{\alpha^{*}})
\end{equation}

and:

\begin{equation}
\label{9.... }
\frac{1}{G^{*}} \approx \frac{1}{G_{N}} (1 - \frac{x}{\beta^{*}})
\end{equation}

and where for $\alpha^{*} $ and $ \beta^{*} $ positive hardening of the microcomposite is present in this order. This is observed in numerical simulation,  \cite{GCL}, where however also crystallographic microstructure is taken into account in the numerical modeling. 

The equations (\ref{7.}) - (\ref{9.}) have different form for the case in which there is no distribution of the elastic moduli of particles of the second material, e.i. when $ G_{0} = G_{1} = G $ and when $ K_{0} = K_{1} = K $. Then we obtain that the equations (\ref{7.}) - (\ref{9.}) take the following form.

The effective shear modulus $G^{*}$ is found from the equation:

\begin{equation}
\label{12.}
\frac{1}{G^{*}} = \frac{1}{G_{N}} - \frac{x}{G_{N} } [\frac{G - G_{N}}{G^{*} + \beta^{*}(G - G^{*})} ]
\end{equation}

and the effective bulk modulus $K^{*}$ is found from the equation:

\begin{equation}
\label{14.}
\frac{1}{K^{*}} = \frac{1}{K_{N}} - \frac{x}{K_{N}} [\frac{K - K_{N}}{K^{*} + \alpha^{*}(K - K^{*})} ].
\end{equation}

In this chapter we have formulated a theory for the elastic moduli of a microcomposite which has two constituents, each of them is isotropic and elastic, the first one has the diameter of particles the same, the second one has a distribution of the form of particles which are almost spheres and which have distribution of their elastic moduli, it´s concentration is $x$. We studied in (\ref{12.}) - (\ref{14.}) the case of no distribution of the form of particles of the second type.

\section{The case of no distribution of the form of particles and effective elastic moduli}

In (\ref{12.}) - (\ref{14.}) we have described the equations for the effective shear and bulk moduli in the case of no distribution of the form of particles of the second type. To study the solution of these equations it is convenient to introduce the following ansatz in the equations (\ref{7.}) and (\ref{9.}):

\begin{equation}
\label{17..}
G^{*} = (G - G_{N}).g \\
\end{equation}
\[ K^{*} = (K - K_{N}).k \]

We have found that the functions g and k are given by:

\begin{equation}
\label{18..}
g_{\pm} = \frac{1}{2[1 - \beta^{*}]} [- \beta^{*}(\gamma + \gamma_{N})  +  (x + \gamma_{N}) \pm \sqrt{D_{g}}]
\end{equation}

where:

\begin{equation}
\label{19..}
D_{g} = [\beta^{*}(\gamma  + \gamma_{N})  - (x + \gamma_{N})]^{2} + 4 (1 - \beta^{*})\beta^{*} \gamma \gamma_{N}
\end{equation}

and where:

\begin{equation}
\label{20.}
\gamma_{N} = \frac{G_{N}}{G - G_{N}}
\end{equation}

and:

\begin{equation}
\label{21.}
\gamma = \frac{G}{G - G_{N}},
\end{equation}
 
and:

\begin{equation}
\label{22.}
k_{\pm} = \frac{1}{2[1 - \alpha^{*}]} [- \alpha^{*}(\delta + \delta_{N})  +  (x + \delta_{N}) \pm \sqrt{D_{k}}]
\end{equation}

where:

\begin{equation}
\label{23.}
D_{k} = [\alpha^{*}(\delta  + \delta_{N})  - (x + \delta_{N})]^{2} + 4 (1 - \alpha^{*})\alpha^{*} \delta \delta_{N}
\end{equation}

and where:

\begin{equation}
\label{24.}
\delta_{N} = \frac{K_{N}}{K - K_{N}}
\end{equation}

and:

\begin{equation}
\label{25.}
\delta = \frac{K}{K - K_{N}}.
\end{equation}

The effective Poison's ratio $\nu^{*}$ has the form:

\begin{equation}
\label{26.}
\nu^{*} = \frac{3k - 2g \epsilon}{6k + 2g \epsilon}
\end{equation}

where $\epsilon $ is defined as:

\begin{equation}
\label{27.}
\epsilon = \frac{G - G_{N}}{K - K_{N}}.
\end{equation}

Note that $D_{g}$ and $D_{k}$ are discriminants which may be positive or negative. In the first case $g_{\pm}$ is real and $k_{\pm}$ is real too.
In the second case $g_{\pm}$ is a  number  and $k_{\pm}$ is a complex number too. Note also that the signs of both discriminants in equations (\ref{18.}) and (\ref{20.}) are independent.
Different signs may be valid in different regions of concentration. Here we will consider the static case only in the following. According to the fluctuation-dissipation theorem the imaginary part of the moduli will be zero. Neverthless note that the results above are obtained for general moduli and that complex moduli can be measured.

For isotropic elastic solids \cite{TG} the range of $\nu^{*} $ of stability is between -1 and 0.5. We will assume that the microcomposite is an isotropic elastic solid. Thus if a percolation transition in this microcomposite occurs at $ \nu^{*} > 0.5 $ it is behind the stability region. Then increasing the concentration of the second component /or decreasing concentration of the second component if this component forms a matrix/ may lead to an unstable microcomposite. The microcomposite is assumed to be unconstrained. Such a results would predict, if we assume that the isotropic elastic solid /microcomposite/ remains isotropic changing the concentration, that it may become mechanically unstable. However let us note that in \cite{TG}  the composite as a continuum isotropic is considered, and this may be not true for large clusters.

The range of the Poison's ration $ \nu^{*} $ above corresponds to the stability criteria, e.i. to positive $G^{*}$ and $K^{*}$ mechanical  moduli. Note however that ferroelastic materials may have a negative stiffness, see in \cite{S}. Then the problem with negative stiffness inclusions /clusters/ in a matrix may be considered by our model, this problem is an interesting one but not discussed here.  Note that in \cite{LET} ferroelastic martensites parent to product phase transition was studied. Order parameter which drives this transition is shear deformation. Strain energy dominates morphology, and the transition results in a characteristic lamellar or twinned structure. This then may lead to a nonisotropic  microcomposite, and calculations of the mechanical moduli should take this fact into account. It is not the case in this paper for simplicity. Martensites are proper ferroelastics, and improper ferroelastics are f.e. ferroelectrics, the martensites show weakly first order phase transition near the second order, or second order phase transition and thus Landau theory applies. This fact is used in our paper. Examples of proper ferroelastics are InTl, FePd, NiTi, AuCd. In proper ferroelastics textures form occurs and it does not correspond with the assumption of the isotropic elastic microcomposite. Improper ferroelastics are materials like high-temperature superconductors, and GMR manganites. In \cite{WL} it was studied how composites with inclusions of negative bulk modulus in a matrix behave.  Let us note that complex values of mechanical moduli correspond to viscoelastic materials, in which however $G^{''} $, the imaginary part of the shear modulus, must be positive \cite{CH} if the material should be stable.
The quantity $\tan(\delta)$ is given by the ratio $ \frac{G^{''}}{G^{'}}$, where $ G^{'}$ is the real part of the shear modulus. Experimental results for this ratio are from 0.01 at room temperature  to 0.015 at higher temperatures. In \cite{JL} it was found that riples are present in the response for temperature dependence of the 5-percent  $VO_{2} $ in Sn matrix.

The shear stress induces large changes in strain when the shear modulus is small in its value. At the percolation transition when one cluster of the dielectric material becomes infinitely large cluster, the mechanical properties of the microcomposite may be expected to  be such that small stress induces large changes in the strain, even that at this transition and near this transition an instability occurs. The material becomes very susceptible to the stress, or even mechanically unstable. While pressure induces changes in clusters of both types of materials in the microcomposite, and thus the susceptibility of the microcomposite to pressure will be present too, the clusters change their volume while interacting, the shear stress leads to shear stress of clusters acting on other clusters. The clusters do not have a regular shape. Due to this fact we may expect that pressure on the microcomposite leads to shear stresses between clusters inside the material. And the shear stress acting on the microcomposite leads to volume changes of clusters induced by other clusters, on which the shear stress is acting. Thus at and near the percolation transition we may expect that both mechanical moduli become  /almost/ zero.

We have found that expansion in $ \beta^{*}$ is useful. When $\beta^{*}$ tends to zero then $G^{*}$ tends to small values. This is found to occur for $ \nu^{*} = \frac{4}{5}$, $\alpha^{*}$ is finite. The microcomposite is however unstable under these conditions.  Zero $\beta^{*}$ would correspond to a critical concentration $x_{c} $ which may be calculated. This is not so unnatural because the clusters change their size going with the concentration to the percolation threshold, thus volume properties of the components are changing depending on their mechanical properties. Note however that the $ \nu^{*} = \frac{4}{5}$ value is beyond the stability region of the material, as noted above. So there exists another critical value of the concentration, $x_{u1}$, above which the composites become mechanically unstable depending on their material constants. We described these results qualitatively, it is not the aim of this paper to study mechanical instability of microcomposites which may appear for certain range of material constants.

\section{Low concentration of the dielectric material: elastic moduli}

From the equation (\ref{12.}) - (\ref{14.}) we obtain in the limit of small concentrations x up to $O(x)$ of the dielectric particles the effective Poison's ratio:

\begin{equation}
\label{17.}
\nu^{*} = \frac{3 K_{N} -  2 G_{N}}{6K_{N} + 2 G_{N}}(1 - x \theta)
\end{equation}

where the constant $ \theta $ is given by:

\begin{equation}
\label{17....}
\theta \equiv (\frac{\frac{3K_{N}}{\delta_{N} + \alpha^{*} \delta} - \frac{2G_{N}}{\gamma_{N} + \beta^{*} \gamma}}{3 K_{N} -  2 G_{N}} - \frac{\frac{6K_{N}}{\delta_{N} + \alpha^{*} \delta} + \frac{2G_{N}}{\gamma_{N} + \beta^{*} \gamma}}{6K_{N} + 2 G_{N}})
\end{equation}

and the effective bulk modulus:

\begin{equation}
\label{18.}
K^{*} = K_{N}(1  - \frac{x}{\delta_{N} + \alpha^{*} \delta})
\end{equation}

and the effective shear modulus:

\begin{equation}
\label{19.}
G^{*} = G_{N}(1 - \frac{x}{\gamma_{N} + \beta^{*} \gamma})
\end{equation}

where $\alpha$ and $\beta$ constants are calculated as above for the material M.

We have found that:

\begin{equation}
\label{19...}
\beta^{*} = \beta_{N}(1 + x ( \frac{5 \nu_{N}\theta}{2 (4 - 5\nu_{N})} -  \frac{ \nu_{N}\theta}{15 (1 - \nu_{N})}))
\end{equation}

and for $ \alpha^{*}$ we have found:

\begin{equation}
\label{19....}
\alpha^{*} = \alpha_{N}(1 - x ( \frac{ \nu_{N}\theta}{3 (1 + \nu_{N})} +  \frac{ \nu_{N}\theta}{3 (1 - \nu_{N})}))
\end{equation}

Note however that in the equations (\ref{18.}) - (\ref{19.}) we can omit the star at the quantities $\alpha$ and $\beta$ due to the fact that the calculations are done up to the second order in the concentration x, i.e. up to the quantities of the order $O(x^{2})$.

From the equations (\ref{18.}) we can find that the material with inclusions is less harder if the microcomposites is such that the quantity $ \delta_{N} + \alpha_{N}\delta $ is positive. And vice versa. From the equation  (\ref{19.}) we can find that the material with inclusions stiffens if the microcomposite is such that the inequality $ \gamma_{N} + \alpha_{N}\gamma $ is positive. And vice versa.

\section{Low concentration of the dielectric material: dielectric response}
  
For low concentration of dielectric particles in the ferroelastic matrix
 we may calculate the dielectric response of the
microcomposite as in \cite{HS}. For values of small $ x $ 
the effective dielectric
permitivity $ \epsilon_{eff}$ is given by:

\begin{equation}
\label{13...}
\epsilon_{eff} = \frac{1}{\alpha{f}}(1 + \frac{ 6 \Gamma p}{\alpha_{f} G_{N} }) + 
\end{equation}
\[ + x \frac{1}{\alpha_{f}}\frac{6 \Gamma p G_{N} }{\delta_{N} + \alpha_{N} \delta}+ 3 x \frac{1}{\alpha_{f}}(1 + \frac{ 6 \Gamma p}{\alpha_{f} G_{N} }).
\frac{\epsilon_{d} - \frac{1}{\alpha_{f}}(1 + \frac{ 6 \Gamma p}{\alpha_{f} G_{N}})}{
\epsilon_{d} + 2 \frac{1}{\alpha_{f}}(1 + \frac{ 6 \Gamma p}{\alpha_{f} G_{N} })} \]

where $ \alpha_{f}$ is the dielectric permitivity of the ferroelastic matrix. We see from the equation (\ref{13...}) that when the material elastically softens due to increasing concentration of dielectric material, i.e. when the conditions for softening described above are fulfilled, then the response of the ferroelastic-dielectric microcomposites is higher. This is true for small concentrations of the dielectric material.

\section{Summary}

We studied here dielectric and mechanical responses of the microcomposites:
ferroelastic-dielectric. A model for such a microcomposite was
formulated in \cite{HS}. Cubic symmetry of the material was assumed which corresponds
to most of the known ferroelastic materials at high temperatures. Nevertheless our results give a qualitative picture also for other symmetries. In our paper \cite{HS} we did not consider the mechanical
inclusion-matrix interactions. This has influence on the mechanical moduli of the composite. The ferroelastic crystals
 as ferroelastic particles are assumed here to have no domain wall structure. We have described a mechanical response of the composite which consists of the material M and of the other material with dispersion of mechanic moduli of particles due to near sphere shape and due to anisotropy. Larger dispersion of mechanical moduli of the second material leads to softening and the stiffness has lower values for small concentrations. For no dispersion case we have found a known result for the effective moduli of the microcomposite. The inclusions of the dielectric material in the ferroelastic matrix have the effect on the dielectric response of the microcomposite. We have found under which conditions the inclusions enhance the dielectric response. Effect of elastic clamping was not discussed in this paper. For its discussion for improper and
 pseudoproper ferroelastic inclusions see \cite{PS}.

Results of low frequency dielectric constant and dielectric loss on
orthorhombic $ Al_{2}(WO_{4})_{3}$ show for polycrystalline material \cite{MVATG},
where voids may play the role of dielectric material, linear
increasing dependence with hydrostatic pressure. This is qualitatively in
agreement with our theory. The authors of these measurements observed hysteresis with increasing and decreasing pressure. This hysteresis may be qualitatively understood within our model due to amorphization of the polycrystalline material at higher pressures and due to the fact that the amorphization remains decreasing then the pressure. Amorphous microcomposite has different mechanical properties than polycrystalline. From our theory it follows that for different mechanical constant values there are different slopes and dielectric response dependence on the pressure. In epitaxial films proper ferroelastic
phase transitions with symmetry-conserving and symmetry-breaking misfit
strains may be present \cite{BL1}. Most of microcomposites are in the form of
films. The authors \cite{BL1} have found that if the extrinsic misfit
strain   does not break the symmetry of the high-temperature phase,
the transition in the film
occurs at somewhat lower temperature than in the bulk. This may play the role in the observation of the transition in this material.

 In cubic-tetragonal systems like $Nb_{3}Sn$,
$V_{3}Si$, $In-Tl$ alloys, $Fe-Pd$ alloys and $ Ni_{2}MnGa$ the cubic
cell elongates (or contracts) in one of the main axis to form tetragonal
cell \cite{CJ}. While these interesting materials are in general
different from those about which we wrote in the Introduction and above,
their high-temperature phase is cubic and our approach may be used to
study them.

\section*{Acknowledgements}

One of the authors (O.H.) wishes to express his sincere thanks to 
Prof. W. Schranz 
for his kind hospitality during his stay in Vienna, and to the University of Vienna, Institute of Experimental Physics, Faculty of Physics, for the financial support of this stay. 
This paper was performed in the frame of the OEAD-WTZ Austria-Czech Republic cooperation project No. 2004/24.

\end{document}